\documentclass[12pt,a4paper]{article}

\usepackage{epsfig}
\usepackage{amsmath,amssymb,amsfonts}
\usepackage{float,a4wide}
\usepackage{cite}
\usepackage{bbm}

\newcommand{\be}{\begin{equation}}
\newcommand{\ee}{\end{equation}}
\newcommand{\bi}{\begin{itemize}}
\newcommand{\ei}{\end{itemize}}
\newcommand{\bea}{\begin{eqnarray}}
\newcommand{\eea}{\end{eqnarray}}
\newcommand{\Det}{\hbox{Det}}
\newcommand{\hbo}{\hbox to 1 true cm {\hfill } }
\newcommand{\tr}{\hbox{tr}}

\def\dslash{\partial\kern-.5em\slash}
\def\kslash{k\kern-.5em\slash}
\def\pslash{p\kern-.5em\slash}
\def\Dslash{D\kern-.5em\slash}

\newcommand{\ket}[1]{|\,#1\,\rangle}          
\newcommand{\linv}{{\triangle}^{-1}}      
\newcommand{\ud}{\mathrm{d}}
\newcommand{\pathD}{\mathcal{D}}
\newcommand{\e}{\mathrm{e}}		
\newcommand{\x}{\boldsymbol{x}}		
\newcommand{\y}{\boldsymbol{y}}		
\newcommand{\z}{\boldsymbol{z}}		
\newcommand{\vv}{\boldsymbol{v}}		
\title{The ice-limit of Coulomb gauge Yang-Mills theory}
\author{T.~Heinzl, A.~Ilderton, K.~Langfeld, M.~Lavelle, D.~McMullan \\ \\
School of Mathematics \& Statistics, University of Plymouth \\
Plymouth, PL4 8AA, UK \\
email: kurt.langfeld@plymouth.ac.uk \\
}
\date{\today}

\begin{document}
\maketitle

\begin{abstract}
In this paper we describe gauge invariant multi-quark states generalising the path integral framework developed by Parrinello, Jona-Lasinio and Zwanziger to amend the
Faddeev-Popov approach. This allows us to produce states such that, in a limit which we call the ice-limit, fermions are dressed with glue exclusively from the fundamental
modular region associated with Coulomb gauge.  The limit can be taken analytically without difficulties, avoiding the Gribov problem. This is illustrated by an unambiguous construction of gauge invariant mesonic states for which we simulate the static quark--antiquark potential.
\end{abstract}

\newpage
\tableofcontents

\section{Introduction}

Yang-Mills theories are the cornerstone of the standard model. Their success is largely based upon perturbation theory where gauge fixing, implemented using the Faddeev-Popov
method \cite{Faddeev:1967fc}, plays a key role. However, Gribov~\cite{Gribov:1977wm} has pointed out that, at a non-perturbative level, the Faddeev-Popov method fails since there
are always gauge equivalent (Gribov) copies which satisfy a chosen gauge condition. Initially this was shown for Coulomb gauge, but Singer has proven that this is in fact a
general problem~\cite{Singer:1978dk}.

There have been various attempts to extend the Faddeev-Popov method to the non-perturbative regime. For example, it was  suggested that the various Gribov copies, weighted by the Faddeev-Popov determinant, should contribute to the functional integral with alternating signs. This approach can be viewed as the insertion of a
topological invariant into the partition function. Unfortunately, that topological invariant turns out to be zero in SU($N_c$) Yang-Mills theory leaving us with the disastrous
conclusion that the generalised Faddeev-Popov method results in physical observables being in indeterminate form~\cite{Neuberger:1986xz, Baulieu:1996kb,Baulieu:1996rp}.

This state of affairs is unfortunate as the need for non-perturbative gauge fixing is widely recognised as physically desirable. For example, Dyson-Schwinger equations are widely
used in hadron phenomenology, and their construction relies on unambiguous gauge fixing, in particular in the infra-red regime.  Using stochastic quantisation to by-pass the
Gribov problem~\cite{Zwanziger:1981kg,Baulieu:1981ec,Horibe:1983ts}, Zwanziger showed~\cite{Zwanziger:2003cf} that the tower of Dyson-Schwinger equations is unchanged but
supplemented with additional constraints reflecting that gauge configurations are confined to the first Gribov region. It turns out that the Green's functions solving the
Dyson-Schwinger equations~\cite{vonSmekal:1997vx,vonSmekal:1997is,Alkofer:2000wg} appear to agree to a large extent with lattice
simulations~\cite{Suman:1995zg,Cucchieri:1997dx,Bloch:2003sk}. We note, however, that some initial discrepancies~\cite{Leinweber:1998uu,Bonnet:2001uh,Langfeld:2001cz} in the
infra-red~\cite{Fischer:2002eq} behaviour of Green's functions have been confirmed in large volume simulations~\cite{Bogolubsky:2007ud,Cucchieri:2008fc}. It became clear only
recently that these findings can be accommodated by the Gribov-Zwanziger approach when the Gribov-Zwanziger action is appropriately modified while preserving renormalisability
and BRST invariance~\cite{Dudal:2007cw,Dudal:2008sp}.

To go beyond the Faddeev-Popov method, we will here use an alternative construction of the partition function~\cite{Parrinello:1990pm,Zwanziger:1990tn} which defines a gauge
invariant action by integrating a weight function over the gauge orbit. This  method has been studied on the lattice in the strong coupling
expansion~\cite{Fachin:1991pu,Parrinello:1991xk}, and in numerical simulations of the weak-coupling regime~\cite{Henty:1996kv}. The phase diagram was explored
in~\cite{Mitrjushkin:2001hr}, and a phase transition from the weak to the strong gauge fixing regime was reported. Finally, it was argued in~\cite{Aubin:2004av} that the gluon
propagator displays gluon confinement. As we shall show, in a particular limit, which we call the ice-limit, the weight function constrains the gauge configurations to unique
representatives of each gauge orbit. Altogether, these form what is called  the fundamental modular region.

In a Hamiltonian framework integration over the gauge group may be used to
define projection operators onto the different non-Abelian charge
(superselection) sectors of the Yang-Mills Hilbert space in the presence of
external charges. This was first emphasised by Polyakov \cite{Polyakov:1978vu}
and Susskind \cite{Susskind:1979up} and subsequently worked out in detail by a
number of authors (see e.g.\ \cite{Gross:1980br,Marchesini:1981kt}). More
recently Zarembo has used this approach to discuss the Yang-Mills mass gap,
confinement and the interquark potential \cite{Zarembo:1998xq,Zarembo:1998qm}
(see also~\cite{Klauder:1996nx,Klauder:1999gd}). A thorough study of U(1)
quantum mechanics along these lines may be found in \cite{Villanueva:1999st}.

In this paper, we introduce gauge invariant external fields (such as
heavy quarks) through the projection techniques
\cite{Polyakov:1978vu,Susskind:1979up,Gross:1980br,Marchesini:1981kt,Zarembo:1998xq,Zarembo:1998qm}
into  the above alternative construction of the partition
function~\cite{Parrinello:1990pm,Zwanziger:1990tn}.
Using lattice regularisation, we will study the ice-limit where
the fields are restricted to the fundamental modular region of Coulomb gauge.
As an illustration, we will  calculate
the static heavy quark--antiquark potential.

\section{Non-perturbative gauge fixing  \label{sec:sol}}

\subsection{The Gribov problem}

Recall that gauge fixing amounts to identifying the space
$\mathfrak{A}/\mathfrak{G}$ of gauge inequivalent configurations (that is the
space $\mathfrak{A}$ of all configurations $A$ modulo gauge transformations $g
\in \mathfrak{G}$) with a subset of the total configuration space:
$\mathfrak{A}/\mathfrak{G} \subset \mathfrak{A}$. The original idea of gauge
fixing (due to Weyl, see \cite{Jackson:2001ia} for the historical context)
attempts at choosing a gauge `slice' $\Gamma$ (of configurations satisfying the
gauge condition) to be identified with the physical configuration space. While
this works for the Abelian case, it fails for the non-Abelian theory due to the
existence of residual gauge copies, as shown by Gribov \cite{Gribov:1977wm},
who was also the first to suggest a possible solution. As the copies only
appear as one moves away from the perturbative small field regime and reaches
what is called the `Gribov horizon' it seems appropriate to just stay within
its interior, i.e.\ within the Gribov region. Mathematically, this is
defined as that neighbourhood of the classical vacuum ($A=0$) where the
Faddeev-Popov operator has a positive spectrum. It turns out,
however, that this `off-limits' prescription is not sufficient. Let us briefly
recapitulate the problem and its (formal) solution.

Following `t~Hooft \cite{tHooft:1981ht} one may formulate the gauge fixing
procedure in terms of distance functionals $S_\mathrm{fix} [A^g] \equiv \| A^g
\|^2$ with $\|\; . \; \|^2$ an appropriate $L^2$ norm. As shown by
Semenov-Tyan-Shanskii and Franke \cite{Semenov-Tyan-Shanskii:1982} as well as
dell'Antonio and Zwanziger  \cite{Zwanziger:1982na} the extrema of
$S_\mathrm{fix}$ define the gauge condition while its Hessian (at the critical
points) is the Faddeev-Popov operator such that the Gribov region
is the domain of positive curvature containing $A=0$, known to be convex and to cover
all orbits \cite{Semenov-Tyan-Shanskii:1982,Zwanziger:1982na}. Its boundary
 is the Gribov horizon where the lowest eigenvalue of the Faddeev-Popov operator
vanishes. These authors also realised that there are copies remaining
within the Gribov region, and one has to restrict configurations even further
to the set $\Lambda$ of global minima,
\be
  \Lambda \equiv \{A \in \mathfrak{A}: S_\mathrm{fix} [A] \le
  S_\mathrm{fix} [A^g], \; \mbox{for all} \; g \in \mathfrak{G} \} \;.
\ee
Note that, by construction, this set is included in the Gribov region and
hence the gauge slice. As pointed out
by van Baal \cite{vanBaal:1991zw} one still requires suitable boundary
identifications within $\partial \Lambda$ endowing the subset $\Lambda \subset
\mathfrak{A}$ with the appropriate topology before it can finally be
identified with the physical configuration space of gauge inequivalent
configurations. In this context the latter is denoted the fundamental modular
region (FMR), see \cite{Fuchs:1994zv,vanBaal:2000zc} for reviews on this
subject. We emphasise at this point that it is gauge invariant by
construction. The details of the embedding $\mathfrak{A}/\mathfrak{G} \subset
\mathfrak{A}$, however, will depend on the gauge fixing (functional) chosen as
its starting point.

\subsection{An alternative implementation of gauge fixing \label{sec:alt}}

As we have discussed, a gauge fixing condition $\chi_a[A]=0$ can always be identified as a stationary point of a  gauge fixing functional
$S_\mathrm{fix}[A^g ]$ such that
\be
\frac{ \delta }{ \delta \theta ^a (x) } \, S_\mathrm{fix}[A^g]
= 0 \implies \chi_a[A]=0\;,
\label{eq:s0}
\ee
where the fields $\theta ^a(x)$ parameterise  the gauge transformation, $g(x) = \exp\{ i \theta ^a (x) \, t^a\}$, with $t^a$ being the generator of the SU($N_c$) gauge group. The
Faddeev-Popov approach is then based on the usual assumption \cite[Chapter 16]{Peskin:1995ev} that one can write 1 as
\be
1 = \int\!\pathD\theta\ \delta(
\chi_a[A^g])\ \Det \bigg(\frac{\delta \chi[A^g]}{\delta \theta}\bigg) \;
\label{eq:s1}\;.
\ee
This, though, is not true non-perturbatively (see Appendix~\ref{sec:a} for more details) as, in fact, the right hand side of (\ref{eq:s1}) is zero.
We will therefore use a different approach
here~\cite{Parrinello:1990pm,Zwanziger:1990tn}, and, after reviewing it, we will
study a series of examples.

The starting point of this approach is the definition of a
gauge invariant effective action $S_\mathrm{eff}[A]$ derived from the gauge fixing functional via the identity
\be
1 =\mathrm{e}^{-S_\mathrm{eff}[A]}
\int {\cal D} g  \; \mathrm{e}^{ S_\mathrm{fix}[A^g] } ,
\label{eq:a1}
\ee
where ${\cal D}g $ is the Haar measure on the gauge group. As it stands, this is a purely formal definition and one might ask if the right hand side of (\ref{eq:a1}) is genuinely 1. In the
continuum this question is hard to address, but using a lattice regulator it becomes clear that this really is a 1. To this end, we need to translate into a lattice formulation
where the potential $A_\mu(x)$ is replaced by link variables $U_\mu(x)$ which transform under a gauge transformation (now conventionally written as $\Omega(x)$) according to
\be
U_\mu ^\Omega (x) = \Omega (x) U_\mu(x)  \Omega^\dagger (x+a e_\mu) \,.
\label{eq:5}
\ee
Inserting (\ref{eq:a1})
into the Yang-Mills partition function we obtain
\be
Z = \int {\cal D} U_\mu \, {\cal D} \Omega \;
\mathrm{e}^{S_\mathrm{fix}[U^\Omega] } \,
\mathrm{e}^{- S_\mathrm{eff}[U] } \,
\mathrm{e}^{S_\mathrm{YM}[U]} \; .
\label{eq:a3}
\ee
For such a lattice regulated partition function, we may interchange the integration over the links
$U_\mu $ and the gauge transformations $\Omega $:
\bea
Z &=&  \int  {\cal D} \Omega \, {\cal D} U_\mu \;
\mathrm{e}^{S_\mathrm{fix}[U^\Omega] } \,
\mathrm{e}^{-S_\mathrm{eff}[U] } \,
\mathrm{e}^{S_\mathrm{YM}[U]} =
\int  {\cal D} \Omega \, {\cal D} U^\Omega _\mu \;
\mathrm{e}^{S_\mathrm{fix}[U^\Omega] } \,
\mathrm{e}^{- S_\mathrm{eff}[U^\Omega] } \,
\mathrm{e}^{S_\mathrm{YM}[U^\Omega ]}
\nonumber \\
&=&
\left(  \int {\cal D} \Omega \right) \; \int {\cal D} U_\mu \,
\mathrm{e}^{S_\mathrm{fix}[U] } \,
\mathrm{e}^{- S_\mathrm{eff}[U] } \,
\mathrm{e}^{S_\mathrm{YM}[U]} ,
\label{eq:a4}
\eea
from the invariance of the action and the Haar measure.
This means that we have been able to factor out the gauge redundancies
into a volume factor in much the same way as the original Faddeev-Popov trick tried to do. However, as we shall see, this procedure is valid non-perturbatively.

We will now investigate how this construction is used in three examples.

\subsection{Three examples \label{sec:trinity}}

\subsubsection{The Christ-Lee Model \label{sec:christlee}}

The Christ--Lee partition function \cite{Christ:1980ku} is given by the two dimensional integral
\be
Z_{_\text{CL}} = \int d^2x \; \mathrm{e}^{S_{_\mathrm{CL}}(\x)} ,
\label{ZCL}
 \ee
where $S_{_\text{CL}}(\x)$ is a function depending only on $r \equiv
\sqrt{x^2+y^2}$. Gauge transformations are rotations
through an angle $\phi$, which we write
$\x\to\x^{\,\phi}$. Taking, for example,
$S_{_\text{CL}}(\x)=-\x^{\,2}$, one may check that $Z_{_\text{CL}} =
2\pi\times 1/2$, where the $2\pi$ comes from the integral over the angle
$\phi$ and $1/2$ is the `physical' partition function. We will consider the
following gauge fixing functional
\be\label{CL-fix}
	S_\text{fix}[\x^\phi] = - \frac{\kappa}{2} \left(\x^\phi - \vv
\right)^2 \;,
\ee
where $\vv$ is an external ``gauge fixing'' vector. The corresponding
gauge condition
$$
\frac{ \partial  S_\mathrm{fix}[\x^\phi]  }{ \partial \phi }
=0 \implies \vv\cdot\x^{\,\phi+\pi/2} =0\;,
$$
exhibits two Gribov copies: $\x^\phi$ parallel or antiparallel to
$\vv$. The FMR is given by those
vectors $\x$ which (globally) maximise the gauge fixing action. In the
present case, these are all vectors of arbitrary length \emph{parallel} to
$\vv$. Without loss of generality, we choose $\vv=(1,0)$ so that the gauge fixing condition becomes $y=0$. The FMR is then given by the positive $x$--axis (with Gribov copies
appearing
on the negative $x$--axis).

In analogy to (\ref{eq:a1}), we find an effective action
\be
\exp\big(S_\text{eff}[\x]\big) = \int\!\ud\phi\ \exp\big(
S_\mathrm{fix}[\x^\phi] \big) = 2\pi I_0(\kappa r)\,\exp\bigg(
-\frac{\kappa}{2}\,r^2-\frac{\kappa}{2}\bigg)\;,
\label{eq:g3}
\ee
with $I_0$ a modified Bessel function of the first kind. The effective action
is manifestly gauge invariant as it depends only on $r$. We now insert the
associated representation  of unity,
\be\begin{split}
	1 &= \exp\big(-S_\text{eff}[\x]\big)\int\!\ud\phi\ \exp\big(
        S_\mathrm{fix}[\x^\phi] \big) \\
	&=\frac{1}{2\pi\, I_0(\kappa r)}\int\!\ud\phi\ \exp\big(\,\kappa\,
        \x^\phi\cdot\vv\,\big)\;,
\end{split}\ee
into the partition function (\ref{ZCL}) where it follows from the discussion
in Section~\ref{sec:alt} that
\be
Z_{_\text{CL}} =
\left( \int\! \ud\phi \right) \, \int\! \ud^2x \ \exp\big( S_{_\mathrm{CL}}(x)\big)\frac{e^{\,\kappa\, x}}{2\pi\,I_0(\kappa
  r)}\;,
\label{eq:g5}
\ee
using $\x\cdot\vv=x$. For $S_{_\text{CL}}=-|\x|^2$, it is a
straightforward matter to perform the $\ud^2x$ integral in (\ref{eq:g5}) and
show that we recover the expected value of $1/2$. The important point is that
the Gribov problem does not hamper this calculation. This can be made explicit
by considering the limit of large $\kappa$, where the asymptotic behaviour of
the modified Bessel function gives
\be
Z_\text{CL} \simeq \left( \int d\phi \right) \, \int d^2x \;
\mathrm{e}^{S_\mathrm{CL}(x)} \,
\sqrt{ \frac{ \kappa \, r  }{2\pi } } \,
\mathrm{e}^{-\kappa\,(r-x)}\;.
\label{eq:g8}
\ee
Importantly, $\kappa(r-x)>0$ so that the $\kappa$ dependent terms of
(\ref{eq:g8}) are a Gaussian regularisation of the delta function. The support
of the delta function arising in the $\kappa \to \infty$ limit
are those vectors $\x$ for which
$$
 \kappa (r - x) = 0 \qquad \implies \qquad y=0\quad \text{and}\quad x\ge 0.
$$
The condition $y=0$ corresponds to our chosen gauge condition, but the
condition $x\geq0$ restricts us to \emph{only} the FMR,
i.e.\ the Gribov copy at $x<0$ is not seen by the partition
function. Explicitly:
\be
\lim_{\kappa \to\infty} \sqrt{\frac{\kappa\,r}{2\pi}}\ \e^{-\kappa (r-x) } =
|x|\, \theta(x)\,\delta(y)\;,
\label{eq:g9}
\ee
where $\theta $ is the Heaviside step function. Using this in (\ref{eq:g8})
our final result for the $\kappa \to \infty$ limit is:
\be
Z_{_\text{CL}} = \bigg(\int\!\ud\phi\bigg)\int\! \ud^2x\
\mathrm{e}^{S_{\mathrm{CL}}(x)} \,
\theta (x) \; \vert x \vert \;  \delta (y) ,
\label{eq:g17}
\ee
Our approach has not only correctly produced the gauge fixing
constraint in terms of the $\delta $ function and the Faddeev--Popov
determinant $\vert x \vert $, but also the correct ``horizon function'' \cite{Zwanziger:1989mf,Zwanziger:1993dh}
$\theta (x)$, which singles out the FMR to the right of the Gribov horizon at $x=0$.
\begin{figure}[t!]
\begin{center}
\includegraphics[width=12cm]{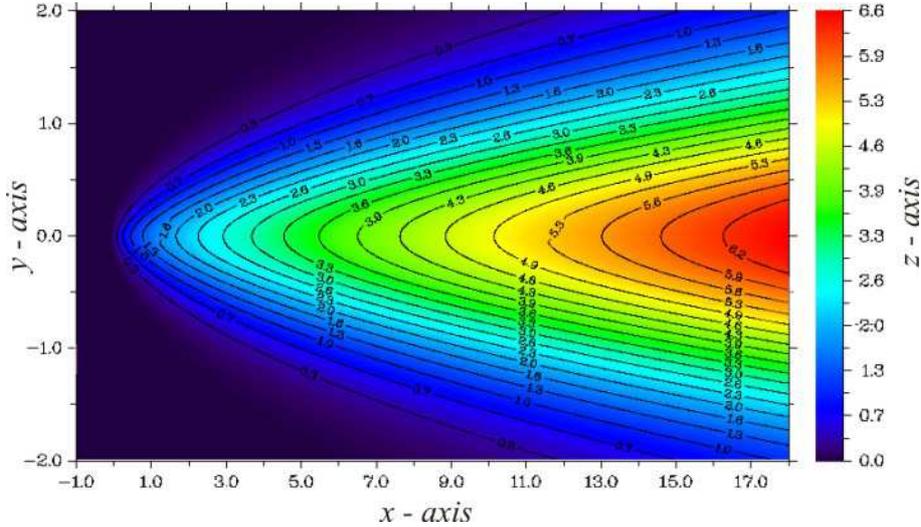}
\end{center}
\caption{Contour plots of the $\kappa$ dependent fraction in (\ref{eq:g5}),
  at $\kappa =16$. The support of this combination of $S_\text{eff}$ and $S_\text{fix}$, and thus of the partition function $Z_{_\text{CL}}$, is restricted to the FMR at large
  $\kappa$.
}
\label{fig:g1}
\end{figure}
In Figure~\ref{fig:g1} we give contour plots of the $\kappa$ dependent fraction in (\ref{eq:g5}), at $\kappa =16$. We clearly see the FMR emerging as the
domain of support. The discussion of a more general class of gauge
fixing functions is left to Appendix~\ref{sec:cl-rev}.

\subsubsection{Example: U(1) Landau gauge}
As a second example we consider U(1) gauge  theory in Landau gauge. Although this does not have a traditional Gribov problem,  zero-modes must be eliminated in order for the
Faddeev-Popov determinant to be non-zero.

The gauge fields $A_\mu (x)$ change under gauge transformations as
\be
\Omega (x) =  \mathrm{e}^{i \lambda (x) } \; , \hbo
A_\mu ^\Omega (x) = A_\mu(x)
+ \partial _\mu \lambda (x) \; .
\label{eq:15}
\ee
We choose Dirichlet boundary
conditions:
\be
A_\mu (x) \vert _{x \in \partial V} = 0 , \hbo
\lambda (x) \vert _{x \in \partial V} = \lambda _0
\label{eq:15b}
\ee
with $\lambda _0$ being constant. The gauge fixing action for Landau gauge
is given by
\be
	S_\text{fix}[A] = -m^2\int\!\ud^4x\ A_\mu A^\mu(x)\;,
\label{eq:15c}
\ee
where the ``mass'' $m$ acts as a gauge fixing parameter. This gauge fixing
action generates the Landau gauge condition via
\be
	\frac{\delta}{\delta\lambda(x)}S_\text{fix}[A^\lambda] =0
\implies \partial^\mu A^\lambda_\mu=0\;.
\ee
Given the boundary conditions (\ref{eq:15b}), it follows that any
$\lambda(x)$ may be written
\be\label{U1BCS}
	\lambda(x)=\lambda_0 + {\overline\lambda}(x)\;.
\ee
where $\lambda_0$ is the surface value of $\lambda(x)$ and
$\overline\lambda(x)$ vanishes on $\partial V$ ($\overline\lambda$ may be decomposed as a sum over the non-zero
Fourier modes of the Laplace operator). The measure on the algebra elements $\lambda$ is inherited from the
group, so
that $\pathD\Omega = \ud\lambda_0 \pathD\bar\lambda$. Note that the gauge fixing action (\ref{eq:15c}) is invariant under
constant gauge transformations since
$A\to A+\partial\lambda= A+\partial\overline\lambda$. We must
adopt some prescription to deal with the zero modes and to this end we will include in our measure a delta function
$\delta(\lambda_0)$ which kills the zero mode and makes the  d'Alembertian $\square$ invertible. We are therefore led to propose
the following form for the effective action:
\be
	e^{S_\text{eff}[A]} = \int\!\pathD\overline\lambda\
        \exp\big(S_\text{fix}[A^{\overline\lambda}]\big) =
 \text{det}^{-1/2}(-m^2\square)\exp\big(-S_\text{fix}[A^T]\big)\;.
\ee
The transverse gauge field $A^T$ is defined by
\be
	A^T_\mu = \bigg(\eta_{\mu\nu} -
        \frac{1}{\square}\partial_\mu\partial_\nu\bigg)A^\nu\;,
\ee
and is gauge invariant. This result may be used in (\ref{eq:a4}) to obtain the
partition function of U(1) gauge theory. We evaluate
\be
	S_\text{fix}[A]-S_\text{eff}[A]=-m^2\int\!\ud^4 x\ A_\mu^L A^{L\mu} + \frac{1}{2}\log\text{det}(-m^2\square)\;,
\ee
where the longitudinal part of the field is defined by $A^L\equiv
A-A^T$\;. Inserting our representation of unity into the partition function we
find that for this U(1) theory
\be
	Z = \text{det}^{-1/2}(-m^2\square)\ \int\!\pathD A_\mu\
        \exp\big(-S_\text{YM}[A]-m^2\int\!\ud^4x\ A^L_\mu A^{L\mu}\big)\;.
\ee
Note that the gauge fixing parameter $m$ acts as a mass for the longitudinal
 gauge fields. In the large mass limit, these decouple from the partition function leaving us with transverse fields only.

\subsubsection{Example: SU(2) and weak gauge fixing}

In our final  example we consider SU(2) Yang-Mills theory, in lattice regularisation, with the gauge fixing functional
\be
S_\mathrm{fix}[U^\Omega] = \kappa \sum _{x, \mu } \tr U_\mu ^\Omega (x) , \hbo \mu = 1 \ldots 4\;,
\label{eq:a20}
\ee
which implies (lattice) Landau gauge upon extremisation. When applied to the partition
function $Z$, the (local) maxima of $S_\text{fix}$ give the
dominant  contributions at large $\kappa$. We refer to this as the `strong
gauge fixing' limit.

As we have illustrated with our previous models the use of the effective
action approach is not limited to the large $\kappa$ regime. Here, we take
$\kappa \ll 1$, the `weak gauge fixing' limit, and calculate the effective
action perturbatively in $\kappa$. Expanding the defining equation
(\ref{eq:a1}) and noting that $S_\mathrm{fix}$ is of order $\kappa$,  we find, up to fourth order in $\kappa $,
\bea
S_\mathrm{eff}[U] &=& \frac{1}{2} \left\langle
S^2_\mathrm{fix} \right\rangle \, + \,
\frac{1}{24} \left[
\left\langle S^4_\mathrm{fix} \right\rangle - 3
\left\langle S^2_\mathrm{fix}\right\rangle^2 \right]
+ {\cal O}(\kappa ^6) \; ,
\label{eq:r11} \\
\left\langle S^n_\mathrm{fix} \right\rangle&:=&
\int {\cal D} \Omega \; S^n_\mathrm{fix}[U^\Omega] .
\eea
For SU(2), terms with an odd power of $\kappa$ vanish upon
integration over $\Omega $. The term quadratic in $\kappa$ is independent of
the links since there is no gauge invariant combination of two link variables
apart from a constant:
\be
\left\langle S^2_\mathrm{fix} \right\rangle
= \frac{\kappa^2}{4} N_l ,
\label{eq:r12}
\ee
where $N_l$ is the number of links on the lattice. The calculation
of the fourth order term is tedious but straightforward. We find:
\be
\left\langle S^4_\mathrm{fix} \right\rangle - 3
\left\langle S^2_\mathrm{fix} \right\rangle^2 =
\frac{\kappa^4}{8} N_l + \frac{\kappa^4}{64} \sum _{p} \frac{1}{2} \tr P_{p}[U] ,
\label{eq:r15}
\ee
where the sum extends over all plaquettes, $P_p$.
Inserting (\ref{eq:r15}) and (\ref{eq:r12}) in (\ref{eq:r11}),
we finally obtain:
\be
S_\mathrm{eff}[U] = N_l \left[ \frac{1}{2} \left( \frac{\kappa }{2}
\right)^2 + \frac{1}{12} \left( \frac{\kappa }{2}
\right)^4 \right] +  \frac{1}{4} \left( \frac{\kappa }{2}
\right)^4  \sum _{p} \frac{1}{2} \tr P_{p}[U]
+ {\cal O}(\kappa ^6) .
\label{eq:r16}
\ee
For sufficiently small  $\kappa $, the effect of $S_\mathrm{eff}[U]$
is just to correct the coefficient $\beta $ of the plaquette
in the Wilson action, showing explicitly that $S_\text{eff}[U]$ is gauge invariant. Terms such as the $1\times 2$ Wilson loop appear at order $\kappa^6$.

\section{Non-perturbative dressing\label{sec:dress} }

In this section we will introduce gauge invariant heavy quarks into the
above approach to the partition function. We will adopt a Schr\"odinger
representation, considering gauge invariant states constructed in a single
time slice. Transition amplitudes between such states will be discussed in
section \ref{sec:heavy}. We begin by briefly reviewing the construction of
gauge invariant charges \cite{Lavelle:1995ty}.

A gauge invariant charged state $\ket{Q}$ may be constructed from a fermionic
state $q(\x)\ket{0}$ by `dressing' the latter with an appropriate function
$h[A]$ of the gauge field $A$,
\be
	\ket{Q}:= h[A](\x)\,q(\x)\ket{q}\;.
\label{eq:16}
\ee
From the transformation properties of the fermion, $q(\x)\to \Omega(\x)q(\x)$,
requiring gauge invariance of the state $\ket{Q}$ implies that the dressing
$h[A]$ transforms as
\be
	h^\Omega[A](\x)= h[A](\x)\,\Omega^\dagger(\x)\;.
\label{dress-def}
\ee
Dressings may be constructed through a field--dependent gauge transformation
which rotates a given field $A$ into a gauge $\chi[A]=0$ . The dressing factor
is not unique and should be selected according to its physical properties \cite{Bagan:1999jf,Bagan:1999jk}.
For example, the dressing describing a single static quark in SU($N_c$) comes from
the field dependent rotation into Coulomb gauge defined by
\be\label{coul}
	\partial_i\bigg(h\, A_i\, h^{-1} + ih\,\partial_i h^{-1}\bigg)=0\;.
\ee
From this example it is clear that the above dressing approach is sensitive to the
Gribov problem of Coulomb gauge: although a
gauge invariant charge may be constructed perturbatively from a solution of
(\ref{coul}), Gribov copies imply that the dressing factor is not well defined
non--perturbatively \cite{Ilderton:2007qy}. This offers a route to understanding confinement as this
lack of single observable quarks is in agreement with experiment. The open
dynamical question is how the Gribov ambiguity produces the physical scale of
hadronic multi--quark systems.

\subsection{Dressing by projection}\label{sec:int-dress}

In this section we review the projection, or group integration, method of
\cite{Polyakov:1978vu,Susskind:1979up,Gross:1980br,Marchesini:1981kt,Zarembo:1998xq,Zarembo:1998qm}
to construct gauge invariant states, and
then make the connection with the partition function above. Consider, for illustration, the group integral
\be
h [A](\x) := \int\!\pathD\Omega\ \Omega (\x) \;
\mathrm{e}^{W[A^\Omega ] } \; ,
\label{eq:41}
\ee
with an, \emph{a priori} arbitrary, weight function $\exp(W[A])$. It may be checked that (\ref{eq:41}) obeys the  transformation law (\ref{dress-def}) and therefore gives a
dressing for a single charge. Dressings for multi--fermion
states may be similarly constructed by inserting the appropriate factors of
$\Omega$ or $\Omega^\dagger$ under the group integral. In this paper  we propose the weight functional $W = S_\text{fix}$, which will allow
us to combine the projection approach with the construction of the
gauge fixed partition function discussed in section~2.

We have made two assumptions in writing down these dressings: (i) the group
integration exists (this is certainly the case in, e.g., lattice gauge
theory), and (ii) the group integration does not yield a vanishing
result for $h[A]$.

The latter assumption is more restrictive. Indeed, we will
see below that (ii) is not fulfilled for the Coulomb dressing, and that
(\ref{eq:41}) needs modifications for this important case. In fact we will see that the required changes allow us, importantly, to attribute a global charge to our locally gauge invariant states.

Let us first look at an Abelian example of this.

\subsection{Example: U(1) dressing and zero modes}
In this example we construct a U(1) dressing from the Coulomb gauge fixing
functional. To illustrate the importance of flat directions, i.e.\ invariance
under some class of gauge transformations, in $S_\text{fix}$, and how they are
related to global charge, we will work in a finite cubic volume $V=L^3$, with
boundary $\partial V$ (recall our states are defined in an initial time
slice). We again take Dirichlet boundary conditions on $A_i(\x)$, and $\lambda(\x)$ constant on $\partial V$ so that we may write $\lambda(\x)=\lambda_0+\overline{\lambda}(\x)$,
analogously to (\ref{U1BCS}).

The gauge fixing functional for Coulomb gauge is
\be
S_\text{fix}[A] = -m \int\limits_V\!\ud^3x\ A_i(\x)A_i(\x)\;.
\ee
We would now like to perform the U(1) analogue of the group integration
(\ref{eq:41}) to construct a dressing for a U(1) fermion. Again, the measure is
$\pathD\Omega=\ud\lambda_0\pathD\overline\lambda$. However, $S_\text{fix}[A]$ is invariant under
constant transformations. It therefore does not see the zero mode $\lambda_0$,
as $S_\text{fix}[A^\lambda]=S_\text{fix}[A^{\overline\lambda}]$. It follows
that the integral over $\lambda_0$ vanishes, and therefore so does the
dressing. To avoid this we once again add $\delta(\lambda_0)$ to the group measure and
find
\be\begin{split}
h[A](\x) &= \int\ud\lambda_0 \pathD\overline\lambda\
\delta(\lambda_0)\ \exp\big(\,i \lambda(\x) + S_\text{fix}[A^{\lambda}]\big)\\
&= \int\!\pathD\overline\lambda\ \exp\big(\,i \overline\lambda(\x) +
S_\text{fix}[A^{\overline\lambda}]\big)\\
&= \mathcal{N} \exp\big(S_\text{fix}[A^T]\big)
        \exp\big(\,i\, \linv \partial_i A_i(\x)\big)\;.
\label{U1ok}
\end{split}\ee
Here $\mathcal{N} =\text{det}^{-1/2}_D(-m\nabla^2)
\exp\big[\linv_{xx}/4m\big]$ is a normalisation factor which is IR finite but
UV divergent. The inverse Laplacian $\linv$ is well defined on $A_i$ as the
fields obey Dirichlet boundary conditions. The transverse field $A_i^T$ is
gauge invariant while the longitudinal term $\linv\partial_i A_i$ is
non--invariant. One may check using any of the expressions in (\ref{U1ok})
that the behaviour of $h$ under a gauge transformation $\alpha(\x)=\alpha_0+\overline{\alpha}(\x)$
is
\be
	h[A^\alpha](\x) =
        h[A](\x)\,\e^{-i\, \overline\alpha(\x)}\;.
\ee
We see that excluding the zero modes from the integral (\ref{U1ok}) has given
us a dressing which transforms as in (\ref{dress-def}) under local
transformations, but is insensitive to global transformations. It follows that
the dressed fermion state transforms with a \emph{constant} phase,
\be
	\ket{Q^\alpha}= h[A](\x)e^{-i\overline\alpha(\x)}\
        e^{i\alpha(\x)}q\ket{0} = e^{i\alpha_0}\ \ket{Q} \;,
\ee
which allows us to assign a global charge of one to these states. We now consider the
analogous Coulomb dressed state in SU($N_c$).

\subsection{Example: multi-quark states in SU($N_c$) }
The Coulomb gauge fixing functional in SU($N_c$), using lattice
regularisation, is
\be
S_\mathrm{fix}[U^{\Omega} ] = \kappa \sum _{x,l=1\ldots 3}
\; \frac{1}{N_c} \tr \, U_l^\Omega (x)\; .
\label{eq:43}
\ee
As in the U(1) example above, $S_\text{fix}$ is invariant under constant gauge
transformations $C$ because of the cyclic invariance of the colour trace. The
corresponding dressing obeys
\be
h [U] = \int {\cal D} (\Omega C) \; \Omega\;
\mathrm{e}^{S_\mathrm{fix}[U^{(\Omega C)} ]}  =
\int {\cal D} (\Omega C)   \; (\Omega C)  \;
\mathrm{e}^{S_\mathrm{fix}[U^{\Omega C} ]}\,C^\dagger  =
h [U] \, C^\dagger \; ,
\label{eq:44}
\ee
and therefore vanishes. The difference between this and the U(1) example is
that the measure on SU($N_c$) does not allow us to separate out the constant
gauge transformations {\it and} preserve the dressing property
(\ref{dress-def}).  For the
remainder of the paper we confine ourselves to multi--quark singlet states
which are invariant under global transformations.  Let us concentrate on physically
relevant states such as mesons and baryons adapting the methods of \cite{Marchesini:1981kt}
for our purposes.

For example, the dressing for a baryonic state in SU($3$) would be

\be
h^{(3)}[U]_{ikm} (\z,\y,\x) =
\int {\cal D} \Omega \; \epsilon _{rst} \, \Omega _{ri} (\z) \;
\Omega _{sk} (\y) \; \Omega _{tm} (\x) \;
\mathrm{e}^{S_\mathrm{fix}[U^\Omega ] } \; .
\label{eq:73}
\ee
The baryonic trial state is invariant under global transformations $C$ since
$$
 \epsilon _{rst} \, C _{ri}  \;
C_{sk}  \; C _{tm}  \; = \;
\hbox{det}(C) \, \epsilon _{ikm}\; = \;
\epsilon _{ikm} .
$$
The dressing for a quark--antiquark state appears as a straightforward
generalisation of (\ref{eq:41}),
\be
h^{(2)}[U](\y,\x)
= \int {\cal D} \Omega \; \Omega ^\dagger (\y) \; \Omega (\x) \;
\mathrm{e}^{S_\mathrm{fix}[U^\Omega ] } \; ,
\label{eq:70}
\ee
This dressing transforms homogeneously,
\be
h^{(2)}[U^G](\y,\x) = G(\y) \, h^{(2)}[U](\y,\x) \,
G^\dagger (\x) \; ,
 \label{eq:71}
\ee
and is invariant under constant transformations. The quark--antiquark state
\be
\vert Q \bar{Q} \,\rangle := \bar{q}(\y) \, h^{(2)}[U](\y,\x)
\, q(\x) \, \vert 0 \rangle
\label{eq:72}
\ee
is therefore gauge invariant.

In the strong gauge fixing limit, the dominant contribution to the dressing
comes from the gauge transformation which \emph{globally} maximises the gauge
fixing functional. This is the transformation $\Omega_\text{FMR}[A](\x)$ which
transforms a given $A$ to its gauge equivalent configuration in the FMR, so
for the mesonic states
\be
\vert Q \bar{Q} \,\rangle\sim \bar{q}(\y) \, \Omega^\dagger_\mathrm{FMR}(\y)
\; \Omega _\mathrm{FMR}(\x) \,
 \, q(\x) \, \vert 0 \rangle\ \e^{S_\text{fix}[A]}\bigg|_{\kappa\gg
   1}\;.
\label{eq:72a}
\ee
In this case, the external quarks are dressed by those gauge transformations
which rotate the gauge  field into the FMR. In the following section we use
this result to derive our main finding -- we will be able to carry out the
strong gauge fixing limit analytically \emph{without} constructing the FMR
explicitly.

\section{Applications\label{sec:heavy}}

\subsection{The heavy quark potential  \label{sec:basic}}

We  now use the formalism above to address, in SU(2) gauge theory, the static
potential of a heavy quark--antiquark pair made gauge invariant via
Coulomb dressing. The ground state energy in this channel depends on the quark--antiquark separation
$r$ and  equals the static potential $V(r)$. Using the
``mesonic dressing'' (\ref{eq:70}), we  calculate
the matrix element (where $r = \vert \x-\y\vert$)
\bea
\rho_\kappa (r,T) &=& \frac{  \langle Q \bar{Q} \, \vert _\kappa
\, \mathrm{e}^{-HT} \, \vert Q \bar{Q} \, \rangle _\kappa }{
\langle 0 \vert  \mathrm{e}^{-HT} \vert 0 \rangle } ,
\label{eq:s10} \\
\vert Q \bar{Q} \, \rangle _\kappa &=&
\bar{q}(\y) \, h^{(2)}(\y; \x) \,
q(\x) \, \vert 0 \rangle \; ,
\nonumber
\eea
where $H$ is the Yang-Mills Hamiltonian. Note the dependence of the states on the gauge fixing parameter $\kappa$. For Coulomb dressing,
the gauge fixing functional is separately defined at each time slice and depends on time $t$ through the time dependence of the
background field $U_\mu (x)$:
\be
S_\mathrm{fix}[U](t) = \kappa \sum _{\x,\, l}
\frac{1}{2} \tr \, U_l(\x,t)  .
\label{eq:s11}
\ee
The static quark--antiquark potential $V(r)$ can be extracted from the
large $T$ limit of $\rho (r,T)$ via
\be
\rho_\kappa (r,T) \to \left\vert \langle 2 \vert Q \bar{Q} \,
\rangle _\kappa  \right \vert^2 \; \mathrm{e}^{- V(r) T} \; ,
\label{eq:s10a}
\ee
where $\vert 2 \rangle $ is the true ground state in the
quark--antiquark channel.
Our aim will be to consider the strong gauge fixing limit $\kappa
\to \infty $ for which the quark--antiquark trial state is dressed
with glue from the FMR (see (\ref{eq:72a})).
Given that the static heavy quark propagator from time $0$ to $T$
is proportional to the so-called ``short'' Polyakov line \cite{Heinzl:2007cp}, i.e.,
\be
 P[U](\x,0,T) = \prod _{t\in [0,T]} U_0(\x,t) \; .
\label{eq:s17}
\ee
we finally obtain for $\rho _\kappa $ in (\ref{eq:s10})
\be
\rho_\kappa (r,T) =  \Bigl\langle
\Omega ^\dagger (\x,0) P[U](\x,0,T) \Omega  (\x,T) \,
\Omega ^\dagger(\y,T) P^\dagger[U](\y,0,T) \Omega  (\y,0) \,
\Bigr\rangle _\mathrm{F} ,
\label{eq:s16}
\ee
where the latter expectation value is defined by
\be
\left\langle O \right\rangle _\mathrm{F} =
N^{-1} \int {\cal D} U_\mu \; {\cal D} \Omega \;
O \; \mathrm{e}^{S_\mathrm{YM}[U] + S_\mathrm{F}[U^\Omega] } \; , \quad
N \equiv \int {\cal D} U_\mu \;  \mathrm{e}^{S_\mathrm{YM}[U]  } .
\label{eq:s18}
\ee
Here the integral $\pathD\Omega$ is initially taken over the gauge transformations in the initial and final time slices, defining the states. However, with a properly normalised Haar measure this may be extended to the entire lattice, which is how we are to understand $\pathD\Omega$ here and below. We have also introduced the total gauge fixing action by
\be
S_\mathrm{F}[U] =S_\mathrm{fix}[U](0)+S_\mathrm{fix}[U](T).
\label{eq:s12}
\ee

\subsection{The weak gauge fixing limit}

The gauge fixing action (\ref{eq:s12}) contains the gauge fixing
parameter $\kappa $ which controls the overlap of the
Coulomb dressed quark--antiquark state with the true ground state.
In fact, for $\kappa =0$, our ansatz for the trial state vanishes,
leaving us with a null result for $\rho _\kappa (r,t)$ in (\ref{eq:s10}).
For small but non-vanishing $\kappa $, the overlap is non-zero, and
we should be able to recover the full static potential
in this limit of weak gauge fixing.

Let us consider $\rho _\kappa (r,T)$ in (\ref{eq:s16}) in leading
order of an expansion with respect to $\kappa $. For this purpose,
we will first perform the integration of $\Omega $ using techniques
familiar from the strong coupling expansion of gauge theories.
To leading order, we may expand the gauge fixing action:
\be
\mathrm{e}^{S_\mathrm{F}[U^\Omega] }
\stackrel{\cdot}{=} \prod _{x,\,l} \left( 1 + \frac{\kappa }{2}
\tr \,U^\Omega_l \right) , \hbo x \in (\x,0), \;  (\x,T) ,
\hbo l=1 \ldots 3.
\label{eq:s27b}
\ee
Subsequently, we perform the integration over the gauge transformations
restricting to the minimal power of $\kappa $ for which (\ref{eq:s27b})
produces a non-vanishing result. This technique is standard textbook material \cite{Montvay:1994cy} so that
we only quote our final result,
\be
\rho _\kappa (r,T) \, = \,
\left(\frac{\kappa}{2} \right)^{2(r/a+1)} \, W(r,t) \; ,
\label{eq:s29}
\ee
where $W(r,t)$ is the rectangular Wilson loop of spatial and temporal extent
$r$ and $T$, respectively. The static potential can
be recovered in the standard fashion
\be
W (r,T) \, \approx \, \vert \langle 2 \vert a \rangle
\vert ^2 \, \mathrm{e}^{- V(r) T} \; ,
\label{eq:s31}
\ee
where $\vert a \rangle $ is the axially dressed quark--antiquark trial
state~\cite{Heinzl:2007kx,Heinzl:2008tv}, usually associated with a
chromo-electric string joining the quark and antiquark.

There are two important implications of (\ref{eq:s29}):
(i) the $\kappa $ dependent prefactor does not influence
the static potential $V(r)$ implying that we analytically find the correct
potential for small but non-zero $\kappa$; (ii) the overlap of our gauge
invariant trial state with the true ground state is quite poor, at least in the weak gauge fixing limit:
\be
\vert \langle 2 \vert Q \bar{Q} \rangle _\kappa
\vert ^2 \, = \, \left(\frac{\kappa}{2} \right)^{2(r/a+1)} \,
\vert \langle 2 \vert a \rangle \vert ^2\,,
\label{eq:s32}
\ee
The overlap is even lower than that associated with ``thin'' Wilson lines. For practical calculations
using lattice gauge theory, we will seek values $\kappa $
of order one.

\subsection{Strong gauge fixing and the ice-limit}

In the last subsection we have shown that we can recover the
static quark--antiquark potential in the weak gauge fixing limit
without being hindered by the Gribov problem. We now show that this also holds for strong gauge fixing.
To contrast our approach with the common difficulties previously encountered we will first show how the Gribov ambiguity reappears for
large values of $\kappa $. However, we will then demonstrate how to circumvent the Gribov problem in the dressing approach by
taking the limit $\kappa \to \infty $ in an appropriate and novel fashion.

Consider the matrix element $\rho _\kappa (r,T)$ given by the functional integral (\ref{eq:s16}), and how we might attempt to perform the orbit integrations $\pathD\Omega$ \textit{before}
the integration over link variables, $U_\mu $. In other words, we integrate over $\Omega$
treating the links $U_\mu$ as a fixed background configuration. Hence, in this step, the matrices
$\Omega $ are considered as the fundamental degrees of freedom of a theory with probability weight $\exp \{ S_\mathrm{F}[U^\Omega] \}$. For  SU(2) gauge theory we have
$$
\Omega = \omega _0 + i \vec{\omega } \cdot \vec{\tau} \, , \hbo
 \omega _0^2 + \vec{\omega }\cdot \vec{\omega } = 1 \; ,
$$
identifying these degrees of freedom as 4-dimensional vectors (spins) of unit
length such that the associated partition function possesses a global $O(4)$ symmetry.
These spins interact only with their nearest neighbours. As the
background links provide a nontrivial ``metric'' for these interactions the partition function describes a spin-glass.
In the strong gauge fixing limit, $\kappa \to \infty$, this spin-glass will approach its ground state. However, for generic background fields, there exists a variety of highly
degenerate near ground-states leading to frustration of the system.
Finding the global maximum of $S_\mathrm{F}[U^\Omega]$,
\be
S_\mathrm{F}[U^\Omega  ] \stackrel{\Omega }{\longrightarrow} \hbox{max} .
\label{eq:rr2}
\ee
amounts to identifying the true ground state of this spin-glass.
It is well known that dealing with this problem is extremely costly and beyond the scope of standard numerical techniques such as
importance sampling. Hence, the Gribov problem has reappeared in disguise, as the problem of simulating a spin glass at low temperatures.

\begin{figure}[t]
\begin{center}
\includegraphics[width=10cm]{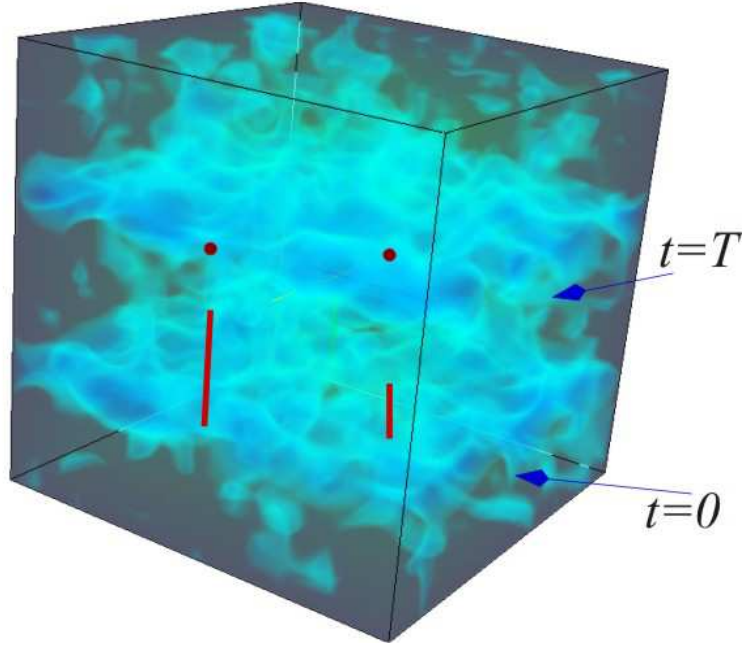}
\end{center}
\caption{ Illustration of the ice-limit: action density and
orientation of the finite length Polyakov lines (vertical lines). The low action planes (at $t=0$ and $t=T$)
are clearly visible.
}
\label{fig:r3}
\end{figure}
The crucial idea to avoid any Gribov (or spin-glass) problem is to \textit{trivialise} the orbit integration over $\Omega$ by factoring it out altogether. This can be done as
follows. First note that in the matrix element $\rho _\kappa $ from (\ref{eq:s16}) we may write
$$
\Omega ^\dagger (\x,0) P[U](\x,0,T) \Omega  (\x,T)
=  P[U^\Omega ](\x,0,T).
$$
Hence, using the gauge invariance of action and Haar measure, we obtain for $N \rho_\kappa$
\bea
&&\int {\cal D} U_\mu \; {\cal D} \Omega \;
\Omega ^\dagger (\x,0) P[U](\x,0,T) \Omega  (\x,T) \,
\Omega ^\dagger(\y,T) P^\dagger[U](\y,0,T) \Omega  (\y,0)
\; \mathrm{e}^{S_\mathrm{YM}[U] +  S_\mathrm{F}[U^\Omega] }
\nonumber \\
&=& \int {\cal D} U_\mu \; {\cal D} \Omega \; P[U^\Omega](\x,0,T)
P^\dagger[U^\Omega](\y,0,T) \; \mathrm{e}^{S_\mathrm{YM}[U]
+  S_\mathrm{F}[U^\Omega] }
\nonumber \\
&=& \int {\cal D} U^\Omega_\mu \; {\cal D} \Omega \; P[U^\Omega](\x,0,T)
P^\dagger[U^\Omega](\y,0,T) \; \mathrm{e}^{S_\mathrm{YM}[U^\Omega]
+  S_\mathrm{F}[U^\Omega] }
\nonumber \\
&=& \Bigl(  \int  {\cal D} \Omega \Bigr) \,
\int {\cal D} U_\mu \;  P[U](\x,0,T)
P^\dagger[U](\y,0,T) \; \mathrm{e}^{S_\mathrm{YM}[U]
+  S_\mathrm{F}[U] } ,
\nonumber
\eea
and, indeed, the gauge group integration factors out as a trivial volume factor, $\int  {\cal D} \Omega = 1$.
Thus, our final answer for $\rho_\kappa$ becomes
\bea
\rho_\kappa (r,T) &=& N^{-1} \,
\int {\cal D} U_\mu \;  P[U](\x,0,T)
P^\dagger[U](\y,0,T) \; \mathrm{e}^{S_\mathrm{YM}[U]
+  S_\mathrm{F}[U] }
\label{eq:s21} \\
N &=& \int {\cal D} U_\mu \;
\mathrm{e}^{S_\mathrm{YM}[U] } .
\label{eq:s22}
\eea
It remains to discuss the strong gauge fixing limit.  In contrast to (\ref{eq:rr2}) we now need to find the maximum of the gauge fixing action \textit{with respect to the links}, $U_\mu $,
\be
S_\mathrm{F}[U] \stackrel{U_\mu}{\longrightarrow } \hbox{max} .
\label{eq:rr11}
\ee
Crucially, this does \textit{not} constitute a spin-glass problem as the global maximum of (\ref{eq:rr11}) is easily obtained:
\be
U_l(\x,t=0) = 1 , \hbo U_l(\x,t=T) = 1 , \hbo
l=1\ldots 3 .
\label{eq:r6}
\ee
Hence, for infinitely large $\kappa$, the spatial links of the time-slices
$t=0$ and $t=T$ are both frozen to the perturbative vacuum.
We therefore call the limit $\kappa \to \infty $ the ``ice-limit''
of the integral dressing approach. The (finite length) Polyakov
lines start and end on the frozen time slices.
Figure~\ref{fig:r3} shows the action density in the cube consisting
of the time axis and two spatial directions.

\begin{figure}[t]
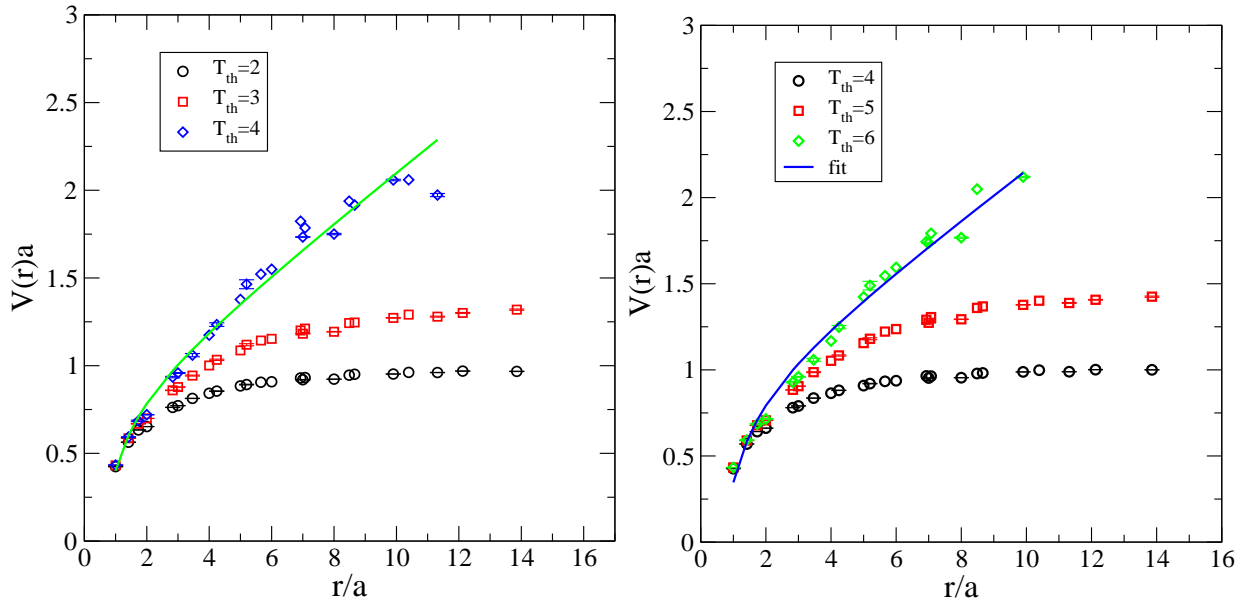

\begin{center}
\includegraphics[width=8cm]{25pot23_2000.eps}
\includegraphics[width=8cm]{25pot23.eps}
\end{center}
\caption{The static quark--antiquark potential from simulations
with $\kappa = 2.0$  (left) and in the ice-limit $\kappa = \infty $
(right).
}
\label{fig:r2}
\end{figure}
We now present our numerical results. For finite values of $\kappa $
(with $\kappa $ not too large to avoid ergodicity problems
in the spin-glass limit), we generated configurations
$(U_\mu, \Omega )$ corresponding to the partition function
\bea
\int {\cal D} U_\mu \; \mathrm{e}^{S_\mathrm{YM}[U] +
S_\mathrm{F}[U] }
&=& \int {\cal D} U_\mu \; {\cal D} \Omega \;
 \mathrm{e}^{S_\mathrm{YM}[U] + S_\mathrm{F}[U^\Omega] }\,,
\label{eq:r1} \\
S_\mathrm{YM}[U] &=& \beta \sum _{x,\mu>\nu} \frac{1}{2} \tr \,
P_{\mu\nu}[U](x) \, ,
\eea
recalling $\int{\cal D}\Omega=1$.
In order to ensure that the excited states are sufficiently suppressed,
the Euclidean time $T$ must be chosen large enough. Here, we have considered
all values $T\ge T_\mathrm{th}$ for the straight-line fit of
\be
- \ln \rho (r,t)_\kappa  =  V(r) \, T \, - \, \ln
\vert \langle 2 \vert Q \bar{Q} \rangle \vert_\kappa  ^2
\label{eq:r4}
\ee
where the asymptotic form  (\ref{eq:s10a}) of $\rho _\kappa (r,t)$
was used.
Our final result for $V(r) a$ using a $16^4$ lattice, $\beta = 2.3$ and
$10000$ independent configurations is shown in Figure~\ref{fig:r2} (left panel).
The line is a fit of the $T_\mathrm{th}=4$ data to
\be
V(r) a = \sigma a^2 \, \frac{r}{a} - \alpha / (r/a) + V_0 ,
\hbo \sigma a^2 (\beta = 2.3) = 0.14 \; ,
\label{eq:r5}
\ee
where the known value for the string tension in lattice units was used.
We find that $T_\mathrm{th}=4$ is sufficient to decouple the excited states.
A good agreement with the potential obtained with standard
overlap enhancing techniques is observed.

We finally study the ice-limit $\kappa \to \infty$.
In this case, configurations were generated using the standard
partition function except that  the spatial links in time slices $t=0$ and $t=T$
were fixed to unity.
Using $\beta =2.3$ and a $16^4$ lattice, the static
potential was extracted from $10,000$ independent configurations
in the ice-limit. The result is shown in Figure~\ref{fig:r2}
(right panel). We observe that rather large values for $T_\mathrm{th}$
are required to decouple the excited states in this case.
One needs $T_\mathrm{th}=6$ to achieve good results for $V(r)$. Note, however, that no
smearing was involved and gauge fixing ambiguities (`Gribov noise' \cite{Heinzl:2007cp}) are absent.

\section{Conclusions}

Non-perturbative gauge fixing is a key ingredient of many approaches to Yang-Mills theory.  In order to have full analytic or numerical control all such approaches must confront
and understand the Gribov problem as this unavoidably arises in any direct imposition of a gauge fixing condition. In this paper we have seen  how to by-pass this problem both in the definition of an unambiguous partition function and in the construction of suitable mesonic states. To this end we have designed a novel and unified framework that combines the properly gauge fixed path integral introduced in~\cite{Parrinello:1990pm,Zwanziger:1990tn} and a generalised dressing approach based on group integration to construct gauge invariant states. The feasibility of this framework has been explicitly demonstrated for various examples ranging from the Christ-Lee model, U(1) gauge theory to SU(2) Yang-Mills theory.

From our new vantage point, the emergence of the Gribov problem could be traced back technically to a problem associated with the order of two integrations extending over the gauge orbits and field configurations, respectively. If the gauge group integration is performed before the average over the gluon fields, the Gribov problem arises as the problem to find the ground state of a spin-glass. However, since gauge invariance is manifest in our approach, the Gribov problem can be avoided by interchanging the integrations over gauge group and gluon fields. In this case, the (`ice') limit of strong gauge fixing can be performed analytically, and the external matter fields become properly dressed by gauge transformations unambiguously connected to the FMR of Coulomb gauge. The numerical feasibility of the method was finally demonstrated by a lattice calculation of the static quark antiquark potential from trial states in the ice limit, living on (initial and final) time slices frozen to the perturbative vacuum.

\vskip 0.5cm
{\bf Acknowledgments:}
The authors thank Andreas Wipf for helpful discussions. The numerical
calculations in this paper were carried out on the HPC and PlymGrid
facilities at the University of Plymouth.

\appendix
\section{Faddeev-Popov method and Gribov problem \label{sec:a}}

The standard approach to (nonabelian) gauge fixing is the
Faddeev-Popov method. Let us briefly recall it in the slightly generalised framework of \cite{Neuberger:1986xz,Baulieu:1996kb,Baulieu:1996rp}. 
In this modified approach one generalises the usual representation of unity (\ref{eq:s1}) to the topological invariant
\bea
N[A] &=& \int {\cal D} \theta^a \; \Det M[A] \; \delta \Bigl(
\frac{ \delta S_\mathrm{fix} [A^\Omega ] }{ \delta
\theta ^a (x) } \Bigr)
\label{eq:s1a} \\
M[A]_{ab}(x,y) &=& \frac{\delta ^2 S_\mathrm{fix} [A^\Omega ] }{ \delta
\theta ^a(x) \delta \theta^b (y) } .
\label{eq:s2a}
\eea
Under the idealising assumption that there is a \textit{unique} solution to the gauge
fixing condition (featuring in the $\delta $-function in (\ref{eq:s1a})) the standard identity (\ref{eq:s1}) is recovered,  
\be
N[A] = 1 \; ,
\label{eq:s3a}
\ee
with a gauge invariant Faddeev-Popov determinant, $\Det M[A] = \Det M[A^g]$.
This may in turn be used to remove the gauge group volume from the partition function,
\bea
Z &=& \int {\cal D} A_\mu \; \mathrm{e}^{S[A]} =
\int {\cal D} A_\mu \; {\cal D} \theta ^b\; \Det M[A] \; \delta \Bigl(
\frac{ \delta S_\mathrm{fix} [A^\Omega ] }{ \delta
\theta^a (x) } \Bigr) \mathrm{e}^{S_\mathrm{YM}[A]}
\nonumber \\
&=& \int {\cal D} A^\Omega_\mu \; {\cal D}\theta ^b \; \Det M[A^\Omega] \;
\delta \Bigl( \frac{ \delta S_\mathrm{fix} [A^\Omega ] }{ \delta
\theta ^a (x) } \Bigr) \;  \mathrm{e}^{S_\mathrm{YM}[A^\Omega]} ,
\nonumber
\eea
exploiting the invariance of the Haar measure, the action and
the Faddeev-Popov determinant. Interchanging the order of integration
and renaming $A^\Omega_\mu \rightarrow A_\mu $ finally yields
\be
Z = \left(\int {\cal D} \theta ^b \right) \;
\int {\cal D} A_\mu \;  \Det M[A] \;
\delta \Bigl( \frac{ \delta S_\mathrm{fix} [A] }{ \delta
\theta ^a (x) } \Bigr) \; \mathrm{e}^{S_\mathrm{YM}[A]} .
\label{eq:s5}
\ee
The latter equation is the desired result: the trivial factor
from the gauge degeneracy has been factored out from the partition
function, and the residual integration can be straightforwardly
evaluated by means of perturbation theory.

The crucial observation, due to Gribov~\cite{Gribov:1977wm},
is that the gauge fixing condition has many solutions. Hence,
the group integration in (\ref{eq:s1a}) becomes a sum over all residual
gauge transformations which cast a given background field $A_\mu $
into one of its Gribov copies. Due to the compactness of the
group integration this implies that (\ref{eq:s3a}) is actually replaced by
\be
N[A] = 0 \; ,
\label{eq:s7}
\ee
so that also this generalised Faddeev-Popov approach remains ill-defined \cite{Baulieu:1996kb,Baulieu:1996rp} .

\section{Christ-Lee model revisited \label{sec:cl-rev}}

In order to make contact with the Faddeev-Popov method we need to evaluate, from (\ref{eq:g5}), the expression
$$
S_\mathrm{eff}[\x] = S_\mathrm{fix}[\x] -  \ln
\int d \phi  \; \mathrm{e}^{ S_\mathrm{fix}[\x^\phi] }
$$
for large values of $\kappa $, and for an arbitrary $S_\text{fix}$. To this end, let $\phi _0$ denote the gauge transformation which transports the vector $\x$ along its orbit to the global maximum of the
gauge fixing action.  For large $\kappa $ we may use a semi-classical approximation to evaluate $S_\text{eff}$, via:
\bea
\ln \int d \phi  \; \mathrm{e}^{ S_\mathrm{fix}[\x^\phi] }  &=&
S_\mathrm{fix}[\x^{\phi _0}] - \frac{1}{2} \ln M[\x]
+ \ln \sqrt{2\pi} ,
\label{eq:g21} \\
M[\x] &=& - \frac{ \partial ^2
S_\mathrm{fix}[\x^\phi] }{ \partial \phi ^2 }
\Bigl\vert _{\phi = \phi _0 } ,
\label{eq:g22}
\eea
where $M$ is the gauge invariant ``Faddeev-Popov matrix''. It follows that (\ref{eq:g5}) becomes
\be
Z = \left( 2\pi \right) \, \int d^2x \; \mathrm{e}^{S_{_\mathrm{CL}}(\x)} \,
\mathrm{e}^{ S_\mathrm{fix}[\x] - S_\mathrm{fix}[\x^{\phi _0}] }
\, \mathrm{e}^{ \frac{1}{2} \ln M[x] - \ln \sqrt{2\pi} }\; .
\label{eq:g23}
\ee
For $S_\mathrm{fix}[\x^\phi]$ as in (\ref{CL-fix})
with $\vv=(1,0)$ and for a given vector $\x=(x,y)$, $\phi_0[\x]$ is defined by
$$
x \, \cos \phi _0 \, - \, y \, \sin \phi _0
\stackrel{\phi_0}{\to } \hbox{max} .
$$
This implies
$$
x \, \cos \phi _0[\x] \, - \, y \, \sin \phi _0 [\x] = r , \hbo
S_\mathrm{fix}[\x^{\phi_0}] = - \frac{\kappa}{2}\big(r-1)^2 .
$$
and indeed the Faddeev-Popov matrix is gauge invariant, $M[\x] = \kappa r$.

Returning to a general $S_\text{fix}$, consider the weight factor $ P(\x) = \exp\{ S_\mathrm{fix}[\x] -
S_\mathrm{fix}[\x^{\phi _0}] \}$.  This will restrict the field (here, $\x$) integration
to the FMR for large values of $\kappa $. To show this, it is convenient to decompose the vector $\x$
into a part $\x_{_\mathrm{FMR}} \in \hbox{FMR}$ and a fluctuation
along the gauge orbit,
$$
\x = \x_{_\mathrm{FMR}}^\varphi .
$$
In contrast to the Faddeev-Popov approach, this must not be done
for the 2-dimensional space as a whole but only locally
for $\x$ close to the FMR. Since
$$
\frac{ \partial  S_\mathrm{fix}[\x^\varphi] }{ \partial \varphi  }
\Bigl\vert _{\x \in \x_{_\mathrm{FMR}} } =  0 ,
$$
we are led to
\be
\mathrm{e}^{ S_\mathrm{fix}[\x] - S_\mathrm{fix}[\x^{\varphi _0}] }
= h(\x_{_\mathrm{FMR}}) \, \delta \left(
\frac{ \partial  S_\mathrm{fix}[\x^\varphi] }{ \partial \varphi  }
\Bigl\vert _{\x \in \x_{_\mathrm{FMR}} } \right) .
\label{eq:g24}
\ee
The weight function $h(\x_{_\mathrm{FMR}})$
can be calculated by integrating $\varphi $ over a small interval around
zero. We will assume that the map $\x \leftrightarrow
(\x_{_\mathrm{FMR}}, \varphi)$ is invertible for $\x $ in a region
around the FMR. We stress, however, that this depends on the gauge
choice and might not be the case for more general settings thus
hinting at a shortcoming of the Faddeev-Popov approach.
Under the above assumption, we use
$$
\int d\varphi \; M(\x_{_\mathrm{FMR}})  \delta \left(
\frac{ \partial  S_\mathrm{fix}[\x^\varphi] }{ \partial \varphi  }
\Bigl\vert _{\x \in \x_{_\mathrm{FMR}} } \right) = 1
\;
$$
and find
\be
h(\x_{_\mathrm{FMR}}) = \int d\varphi \; M(\x_{_\mathrm{FMR}})
\; \mathrm{e}^{ S_\mathrm{fix}[\x] -
S_\mathrm{fix}[\x^{\varphi _0}] }
\stackrel{\cdot }{=} M(\x_{_\mathrm{FMR}}) \int d\varphi \;
\mathrm{e}^{ - \frac{1}{2} M(\x_{_\mathrm{FMR}}) \, \varphi ^2 } =
\sqrt{ 2\pi  M(\x_{_\mathrm{FMR}}) }.
\label{eq:g25}
\ee
Inserting (\ref{eq:g25}) and (\ref{eq:g24}) in (\ref{eq:g23}),
we finally obtain the desired result,
\be
Z = \left( 2\pi \right) \, \int d^2x \; \mathrm{e}^{S_{_\text{CL}}(\x)} \,
M(\x_{_\mathrm{FMR}}) \, \delta \left(
\frac{ \partial  S_\mathrm{fix}[\x^\varphi] }{ \partial \varphi  }
\Bigl\vert _{\x \in \x_{_\mathrm{FMR}} } \right) ,
\label{eq:g26}
\ee
which is the starting point of the Faddeev-Popov method.

\end{document}